\documentstyle[12pt,epsfig]{article}

\newcommand{\bea}{\begin{eqnarray}}
\newcommand{\eea}{\end{eqnarray}}
\begin{document}

\title{Ground state angular momenta of random nuclei} 

\author{R. Bijker$^{1,2}$ and A. Frank$^{2,3}$\\
\mbox{}\\
$^1$ Dipartimento di Fisica, Universit\`a degli Studi di Genova, \\
Via Dodecaneso 33, I-16146 Genova, Italy 
\footnote{Sabattical leave} \\ 
$^2$ Instituto de Ciencias Nucleares, UNAM, AP 70-543, \\ 
04510 M\'exico, DF, M\'exico \\
$^3$Centro de Ciencias F{\'{\i}}sicas, UNAM, AP 139-B, \\ 
62251 Cuernavaca, Morelos, M\'exico}

\date{}

\maketitle

\begin{abstract} 
We investigate the phenomenom of emerging regular spectral 
features from random interactions. In particular, we address the dominance 
of $L=0$ ground states in the context of the vibron model and the 
interacting boson model. A mean-field analysis links different regions of 
the parameter space with definite geometric shapes. The results that 
are, partly, obtained in closed analytic form, provide a clear and 
transparent interpretation of the distribution of ground state angular 
momenta as observed before in numerical studies.   
\end{abstract}

\section{Introduction}

Recent shell model calculations of even-even nuclei in the $sd$ shell 
and the $pf$ shell with random interactions showed a remarkable 
statistical preference for ground states with $L=0$, despite the 
random nature of the two-body matrix elements, both in sign and in relative 
magnitude \cite{JBD}. 
A similar preponderance of $L=0$ ground states was found in an analysis 
of the Interacting Boson Model (IBM) with random interactions \cite{BF1}. 
In addition, in the IBM evidence was found for both vibrational and 
rotational band structures \cite{BF1,BF2}. 
These are surprising results in the sense that, according to the conventional 
ideas in the field, the occurrence of $L=0$ ground states and the existence 
of vibrational and rotational bands are due to very specific forms of the 
interactions. The studies with random interactions show that the class of 
Hamiltonians that lead to these regular features is much larger than usually 
thought. 
These unexpected results have sparked a large number of investigations to 
explain and further explore the properties of random nuclei, both for the 
nuclear shell model \cite{BFP1,JBDT,BFP2,MVZ,DD,DW,ZA,Zuker} and the IBM 
\cite{BF1,BF2,BFP2,KZC,DK,BF3}.   

In this contribution, we investigate the origin of the dominance of 
$L=0$ ground states that emerges from random interactions in two different 
models: the vibron model and the IBM. The vibron model is mathematically 
simpler than the IBM, but exhibits many of the same qualitative 
features. In this case, most of the results can be obtained in closed 
analytic form. For both models, we suggest to use a mean-field analysis 
to address the problem of the probability distribution of the ground state 
angular momenta \cite{BF3}. 

\section{The vibron model} 

The vibron model was introduced to describe the rotational and 
vibrational excitations of diatomic molecules \cite{vibron}. Its building 
blocks are a dipole boson $p^{\dagger}$ with $L^P=1^-$ and a 
scalar boson $s^{\dagger}$ with $L^P=0^+$. The total number of bosons $N$ 
is conserved by the vibron Hamiltonian. We only consider 
one- and two-body interactions 
\bea
H &=& \frac{1}{N} \left[ H_1 + \frac{1}{N-1} H_2 \right] ~, 
\label{h12}
\eea
with 
\bea
H_1 &=& \epsilon_s \, s^{\dagger} \cdot \tilde{s}  
- \epsilon_p \, p^{\dagger} \cdot \tilde{p} ~, 
\nonumber\\
H_2 &=& u_0 \, \frac{1}{2} \, 
(s^{\dagger} \times s^{\dagger})^{(0)} \cdot 
(\tilde{s} \times \tilde{s})^{(0)} 
+ u_1 \, (s^{\dagger} \times p^{\dagger})^{(1)} \cdot 
(\tilde{p} \times \tilde{s})^{(1)} 
\nonumber\\ 
&+& \sum_{\lambda=0,2} c_{\lambda} \, \frac{1}{2} \, 
(p^{\dagger} \times p^{\dagger})^{(\lambda)} \cdot  
(\tilde{p} \times \tilde{p})^{(\lambda)} 
\nonumber\\
&+& v_0 \, \frac{1}{2\sqrt{2}} \, \left[ 
  (p^{\dagger} \times p^{\dagger})^{(0)} \cdot 
  (\tilde{s} \times \tilde{s})^{(0)} 
+ H.c. \right] ~, 
\label{hvib}
\eea
Here $\tilde{s}=s$ and $\tilde{p}_m=(-1)^{1-m}p_{-m}$~. We have scaled 
$H_1$ by $N$ and $H_2$ by $N(N-1)$ in order to remove the $N$ dependence 
of the matrix elements. The seven parameters of the Hamiltonian,  
$\epsilon_s$, $\epsilon_p$, $u_0$, $u_1$, $c_0$, $c_2$, $v_0$,  
altogether denoted by $\vec{x}$, are taken as independent random numbers on 
a Gaussian distribution with zero mean. In this way, the interaction terms 
are arbitrary and equally likely to be attractive or repulsive. 

The connection between the vibron Hamiltonian, potential energy surfaces, 
equilibrium configurations and geometric shapes can be 
investigated by means of mean-field methods \cite{onno}. For the vibron 
model, the coherent state can be written as a condensate of deformed bosons, 
which are superpositions of scalar and dipole bosons 
\bea
\left| \, N,\alpha \, \right> \;=\; \frac{1}{\sqrt{N!}} \, 
\left( \cos \alpha \, s^{\dagger} + \sin \alpha \, p_0^{\dagger} 
\right)^N \, \left| \, 0 \, \right> ~, 
\label{trial}
\eea
with $0 \leq \alpha \leq \pi/2$. The potential energy surface is then given 
by the expectation value of the vibron Hamiltonian of Eqs.~(\ref{h12}) 
and (\ref{hvib}) in the coherent state 
\bea
E(\alpha) &=&  a_4 \, \sin^4 \alpha + a_2 \, \sin^2 \alpha + a_0 ~. 
\label{surface}
\eea
The coefficients $a_i$ are linear combinations of the parameters 
of the Hamiltonian 
\bea
a_4 &=& \vec{r} \cdot \vec{x} \;=\; \frac{1}{2} u_0 + u_1 
+ \frac{1}{6} c_0 + \frac{1}{3} c_2 + \frac{1}{\sqrt{6}} v_0 ~, 
\nonumber\\
a_2 &=& \vec{s} \cdot \vec{x} \;=\; -\epsilon_s + \epsilon_p 
- u_0 - u_1 - \frac{1}{\sqrt{6}} v_0 ~, 
\nonumber\\
a_0 &=& \vec{t} \cdot \vec{x} \;=\; \epsilon_s + \frac{1}{2} u_0 ~. 
\label{coef}
\eea
For random interaction strengths, we expect the trial wave function of 
Eq.~(\ref{trial}) and the energy surface of Eq.~(\ref{surface}) to 
provide information on the distribution of shapes that the model can 
acquire. The value of $\alpha_0$ that characterizes the equilibrium 
configuration of the potential energy surface only depends on the 
coefficients $a_4$ and $a_2$. The $a_2 a_4$ plane 
can be divided into different areas according to the three possible 
equilibrium configurations: $S_1$ for the $s$-boson or spherical 
condensate ($\alpha_0=0$), $S_2$ for the deformed condensate 
($0 < \alpha_0 < \pi/2$), and $S_3$ for the $p$-boson condensate 
($\alpha_0=\pi/2$). The distribution of shapes for an ensemble of  
Hamiltonians depends on the joint probability distribution of the 
coefficients $a_4$ and $a_2$ which is given by a bivariate normal 
distribution 
\bea
P(a_4,a_2) &=& \frac{1}{2\pi \sqrt{|\det M|}} \, 
\mbox{exp} \left[-\frac{1}{2} 
\left( \begin{array}{cc} a_4 & a_2 \end{array} \right) M^{-1} 
\left( \begin{array}{c} a_4 \\ a_2 \end{array} \right) \right] ~, 
\label{bivariate}
\eea
with 
\bea
M &=& \left( \begin{array}{ccc} 
\vec{r} \cdot \vec{r} &\;\;\;& \vec{r} \cdot \vec{s} \\
\vec{r} \cdot \vec{s} && \vec{s} \cdot \vec{s} \end{array} 
\right) \;=\; \frac{1}{18} \left( \begin{array}{rrr} 
28 &\;\;\;& -30 \\ -30 && 75 \end{array} \right) ~. 
\eea
The vectors $\vec{r}$ and $\vec{s}$ are defined in Eq.~(\ref{coef}). 
The probability that the equilibrium shape of an ensemble of Hamiltonians 
is spherical can be obtained by integrating $P(a_4,a_2)$ over 
the appropriate range $S_1$ ($a_2>0$ and $a_4>-a_2$) 
\bea
P_1 &=& \frac{1}{4\pi} \left[ \pi+2 \arctan \sqrt{\frac{27}{16}} \right]  
\;=\; 0.396 ~. 
\label{P1}
\eea
Similarly, the probability for the occurrence of a deformed shape is 
obtained by integrating $P(a_4,a_2)$ over $S_2$ ($-2a_4<a_2<0$)  
\bea
P_2 &=& \frac{1}{2\pi} \arctan \sqrt{\frac{64}{3}} \;=\; 0.216 ~. 
\label{P2}
\eea
Finally, the probability for finding the third solution, a $p$-boson 
condensate, is 
\bea
P_3 \;=\; 1-P_1-P_2 \;=\; 0.388 ~. 
\label{P3}
\eea
The rotational spectrum can be obtained from the angular momentum content 
of the condensate in combination with the Thouless-Valatin formula for 
the moment of inertia \cite{duke}. The ordering of the rotational energy 
levels is determined by the sign of the moment of inertia 
\bea
E_{\rm rot} &=& \frac{1}{2{\cal I}} L(L+1) ~.
\eea

In Table~\ref{percvib} we show the probability distribution of the ground 
state angular momenta as obtained in the mean-field analysis. There is a 
statistical preference for $L=0$ ground states. This is largely due to the 
occurrence of a spherical shape (whose angular momentum content is just 
$L=0$) for almost 40 $\%$ of the cases (see Eq.~(\ref{P1})). The deformed 
shape corresponds to a rotational band with $L=0,1,\ldots,N$, whose ground 
state has $L=0$ for positive values of the moment of inertia ${\cal I}>0$, 
and $L=N$ for ${\cal I}<0$. The third solution, the $p$-boson condensate, 
has angular momenta $L=N,N-2,\ldots,1$ or $0$. For ${\cal I}>0$, the ground 
state has $L=0$ or $L=1$, depending whether the total number of vibrons 
$N$ is even or odd, whereas for ${\cal I}<0$ it has the maximum value 
$L=N$. The sum of the $L=0$ and $L=1$ percentages is constant. 

\begin{table}
\centering
\caption[]{\small 
Percentages of ground states with $L=0$, $1$ and $N$, obtained 
in a mean-field analysis of the vibron model.}
\label{percvib}
\vspace{15pt}
\begin{tabular}{crrrrl}
\hline
& & & & & \\
\multicolumn{2}{c}{Shape} & $L=0$ & $L=1$ & $L=N$ & \\
& & & & & \\
\hline
& & & & & \\
$  \alpha_0=0$       & 39.6 $\,\%$ 
& 39.6 $\,\%$ &  0.0 $\,\%$ &  0.0 $\,\%$ & \\
& & & & & \\
$0< \alpha_0 <\pi/2$ & 21.6 $\,\%$ 
& 13.3 $\,\%$ &  0.0 $\,\%$ &  8.3 $\,\%$ & \\
& & & & & \\
$  \alpha_0=\pi/2$   & 38.8 $\,\%$ 
 & 17.9 $\,\%$ &  0.0 $\,\%$ & 20.9 $\,\%$ & $N=2k$ \\
&&  0.0 $\,\%$ & 17.9 $\,\%$ & 20.9 $\,\%$ & $N=2k+1$ \\
& & & & & \\
\hline
& & & & & \\
Total & 100.0 $\,\%$ & 70.8 $\,\%$ &  0.0 $\,\%$ & 29.2 $\,\%$ & $N=2k$ \\
      &              & 52.9 $\,\%$ & 17.9 $\,\%$ & 29.2 $\,\%$ & $N=2k+1$ \\
& & & & & \\
\hline
\end{tabular}
\end{table}

\begin{figure}
\centerline{\hbox{\epsfig{figure=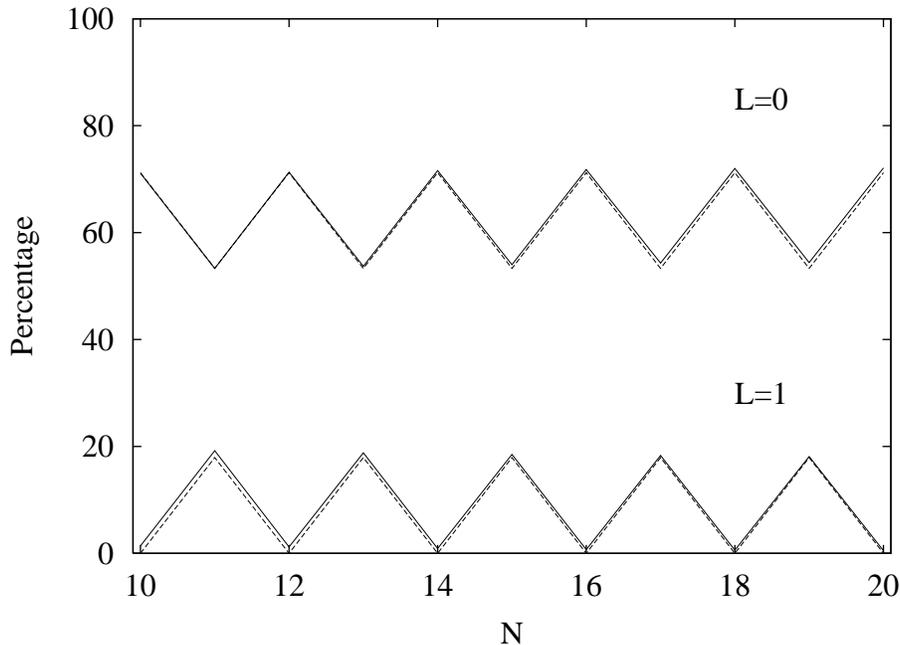} }}
\caption[]{\small 
Percentages of ground states with $L=0$ and $L=1$ in the vibron 
model with random one- and two-body interactions calculated exactly for 
100000 runs (solid lines) and in mean-field approximation (dashed lines).}
\label{vibgs}
\end{figure}

Fig.~\ref{vibgs} shows that the mean-field results for the 
percentages of $L=0$ and $L=1$ ground states 
are in excellent agreement with the exact ones. 
As is clear from the results presented in Table~\ref{percvib}, the 
fluctuations in the percentages of $L=0$ and $L=1$ ground states with $N$ 
are due to the contribution from the $p$-boson condensate solution. 

\section{The interacting boson model}

The interacting boson model (IBM) describes collective excitations in nuclei 
in terms of a system of $N$ interacting bosons \cite{IBM}. Its building 
blocks are a quadrupole boson $d^{\dagger}$ with $L^P=2^+$ and a 
scalar boson $s^{\dagger}$ with $L^P=0^+$. The total number of bosons 
$N$ is conserved by the IBM Hamiltonian. We consider the 
one- and two-body Hamiltonian of Eq.~(\ref{h12}) with 
\bea
H_1 &=& \epsilon_s \, s^{\dagger} \cdot \tilde{s}  
+ \epsilon_d \, d^{\dagger} \cdot \tilde{d} ~, 
\nonumber\\
H_2 &=& \frac{1}{2} \, u_0 \, 
(s^{\dagger} \times s^{\dagger})^{(0)} \cdot 
(\tilde{s} \times \tilde{s})^{(0)} 
+ u_2 \, (s^{\dagger} \times d^{\dagger})^{(2)} \cdot 
(\tilde{d} \times \tilde{s})^{(2)} 
\nonumber\\ 
&&+ \sum_{\lambda=0,2,4} \frac{1}{2} \, c_{\lambda} \, 
(d^{\dagger} \times d^{\dagger})^{(\lambda)} \cdot  
(\tilde{d} \times \tilde{d})^{(\lambda)} 
\nonumber\\
&&+ \frac{1}{2\sqrt{2}} \, v_0 \, \left[ 
  (d^{\dagger} \times d^{\dagger})^{(0)} \cdot 
  (\tilde{s} \times \tilde{s})^{(0)} 
+ H.c. \right]
\nonumber\\
&&+ \frac{1}{2} \, v_2 \, \left[ 
  (d^{\dagger} \times d^{\dagger})^{(2)} \cdot 
  (\tilde{d} \times \tilde{s})^{(2)} 
+ H.c. \right] ~.  
\label{hibm}
\eea
The nine coefficients $\epsilon_s$, $\epsilon_d$, $u_0$, $u_2$, 
$c_0$, $c_2$, $c_4$, $v_0$, $v_2$, are chosen independently from a Gaussian 
distribution of random numbers with zero mean \cite{BF1,BF2}. Just as for 
the vibron model, the connection between the IBM, potential energy surfaces, 
equilibrium configurations and geometric shapes, can be studied with 
mean-field Hartree-Bose techniques by means of coherent states \cite{duke,cs}. 
The coherent state can be written as an axially symmetric condensate 
\bea
\left| \, N,\alpha \, \right> \;=\; \frac{1}{\sqrt{N!}} 
\left( \cos \alpha \, s^{\dagger} + \sin \alpha \, d_0^{\dagger} 
\right)^N \, \left| \, 0 \, \right> ~, 
\eea
with $-\pi/2 < \alpha \leq \pi/2$. The angle $\alpha$ is related to 
the deformation parameters in the intrinsic frame, $\beta$ and $\gamma$ 
\cite{IBM,cs}. The potential energy surface is then 
given by the expectation value of the Hamiltonian in the coherent state 
\bea
E_N(\alpha) \;=\; a_4 \, \sin^4 \alpha + 
a_3 \, \sin^3 \alpha \cos \alpha + a_2 \, \sin^2 \alpha + a_0 ~.  
\eea
For the IBM, the structure of the energy surface 
is a bit more complicated than for the vibron model due to the presence 
of the $a_3$ term. This precludes an analytic treatment as presented 
for the vibron model, but qualitatively the results are very similar. 
In practice, for each Hamiltonian the minimum of the energy surface 
$E(\alpha)$ is determined numerically. Again, the equilibrium configurations 
can be divided into three different classes: an $s$-boson or spherical 
condensate ($\alpha_0=0$), a deformed condensate with prolate 
($0 < \alpha_0 < \pi/2$) or or oblate symmetry ($-\pi/2 < \alpha_0 < 0$), 
and a $d$-boson condensate ($\alpha_0=\pi/2$). 
Each equilibrium configuration has its own characteristic angular momentum 
content. Even though we do not explicitly project the angular momentum 
states from the coherent state, the angular momentum of the ground state 
can, to a good approximation, be obtained from the rotational structure of 
the condensate in 
combination with the Thouless-Valatin formula for the corresponding moments 
of inertia \cite{duke}. The results are summarized in Table~\ref{percibm}. 

\begin{table}
\centering
\caption[]{\small 
Percentages of ground states with $L=0$, $2$ and $2N$, obtained 
in a mean-field analysis of the interacting boson model.}
\label{percibm}
\vspace{15pt}
\begin{tabular}{crrrrl}
\hline
& & & & & \\
\multicolumn{2}{c}{Shape} & $L=0$ & $L=2$ & $L=2N$ & \\
& & & & & \\
\hline
& & & & & \\
$  \alpha_0=0$       & 39.4 $\,\%$ 
& 39.4 $\,\%$ &  0.0 $\,\%$ &  0.0 $\,\%$ & \\
& & & & & \\
$0<|\alpha_0|<\pi/2$ & 36.8 $\,\%$ 
& 23.7 $\,\%$ &  0.0 $\,\%$ & 13.1 $\,\%$ & \\
& & & & & \\
$  \alpha_0=\pi/2$   & 23.8 $\,\%$ 
 & 13.5 $\,\%$ &  0.0 $\,\%$ & 10.3 $\,\%$ & $N=6k$ \\
&&  0.2 $\,\%$ & 13.2 $\,\%$ & 10.4 $\,\%$ & $N=6k+1,6k+5$ \\
&&  4.4 $\,\%$ &  9.0 $\,\%$ & 10.4 $\,\%$ & $N=6k+2,6k+4$ \\
&&  9.3 $\,\%$ &  4.0 $\,\%$ & 10.5 $\,\%$ & $N=6k+3$ \\
& & & & & \\
\hline
& & & & & \\
Total & 100.0 $\,\%$ 
 & 76.6 $\,\%$ &  0.0 $\,\%$ & 23.4 $\,\%$ & $N=6k$ \\
&& 63.3 $\,\%$ & 13.2 $\,\%$ & 23.5 $\,\%$ & $N=6k+1,6k+5$ \\
&& 67.5 $\,\%$ &  9.0 $\,\%$ & 23.5 $\,\%$ & $N=6k+2,6k+4$ \\
&& 72.4 $\,\%$ &  4.0 $\,\%$ & 23.6 $\,\%$ & $N=6k+3$ \\
& & & & & \\
\hline
\end{tabular}
\end{table}

(i) The $s$-boson condensate corresponds to a spherical shape. 
Whenever such a condensate occurs (in 39.4 $\%$ of the cases), 
the ground state has $L=0$.

(ii) The deformed condensate corresponds to an axially symmetric deformed 
rotor. The ordering of the rotational energy levels $L=0,2,\ldots,2N$ is 
determined by the sign of the moment of inertia 
\bea
E_{\rm rot} \;=\;  \frac{1}{2{\cal I}_3} L(L+1) ~. 
\eea
The deformed condensate occurs in 36.8 $\%$ of the cases. For ${\cal I}_3>0$ 
the ground state has $L=0$ (23.7 $\%$), while for ${\cal I}_3<0$ the ground 
state has the maximum value of the angular momentum $L=2N$ (13.1 $\%$). 

(iii) The $d$-boson condensate corresponds to a quadrupole oscillator with 
$N$ quanta. Its rotational structure has a more complicated structure 
than the previous two cases. It is characterized by the labels $\tau$, 
$n_{\Delta}$ and $L$. The boson seniority $\tau$ is given by 
$\tau=3n_{\Delta}+\lambda=N,N-2,\ldots,1$ or 
$0$ for $N$ odd or even, and the values of the angular momenta are 
$L=\lambda,\lambda+1,\ldots,2\lambda-2,2\lambda$ \cite{IBM}. In this case, 
the rotational excitation energies depend on two moments of inertia 
\bea
E_{\rm rot} \;=\; \frac{1}{2{\cal I}_5} \tau(\tau+3) 
+ \frac{1}{2{\cal I}_3} L(L+1) ~. 
\eea
The $d$-boson condensate occurs in 23.8 $\%$ of the cases. 
For ${\cal I}_5>0$ the ground state has $\tau=0$ 
for $N$ even or $\tau=1$ for $N$ odd ($\sim$ 4 $\%$), while for 
${\cal I}_5<0$ the ground state has the maximum value of the boson 
seniority $\tau=N$ ($\sim$ 19 $\%$). For $\tau=0$ and $\tau=1$ 
there is a single angular momentum state with $L=0$ and $L=2$, 
respectively. For  the $\tau=N$ multiplet, the angular momentum of the 
ground state depends on the sign of the moment of inertia ${\cal I}_3$. 
For ${\cal I}_3>0$ the ground state has $L=0$ for $N=3k$ or 
$L=2$ for $N \neq 3k$ ($\sim$ 9 $\%$), while for ${\cal I}_3<0$ the ground 
state has the maximum value of the angular momentum $L=2N$ ($\sim$ 10 $\%$). 

\begin{figure}
\centerline{\hbox{\epsfig{figure=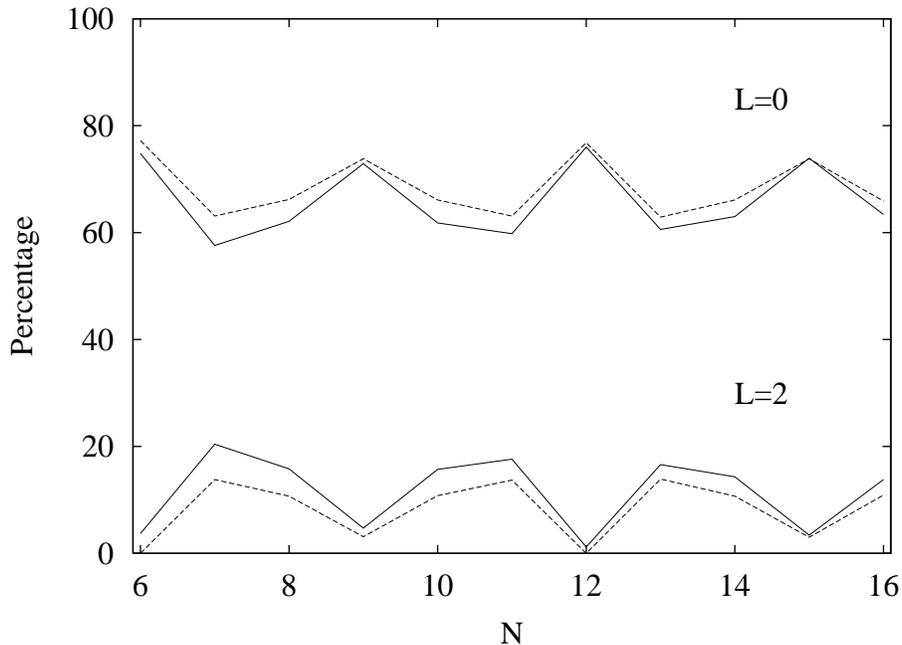} }}
\caption[]{\small 
Percentages of ground states with $L=0$ and $L=2$ in the IBM 
with random one- and two-body interactions calculated exactly for 
10000 runs (solid lines) and in mean-field approximation (dashed lines).}
\label{ibmgs}
\end{figure}

In Fig.~\ref{ibmgs} we show the percentages of ground states with 
$L=0$ and $L=2$ as a function of the total number of bosons $N$. 
A comparison of the results of the mean-field analysis (dashed lines) 
and the exact ones (solid lines) shows a good agreement. There is a 
dominance of ground states with $L=0$. The oscillations of the $L=0$ and 
$L=2$ percentages with $N$ are due to the contribution of the $d$-boson 
condensate (see Table~\ref{percibm}). For $N=3k$ we see an enhancement 
for $L=0$ and a corresponding decrease for $L=2$. In the mean-field 
analysis, the sum of the two hardly depends on the number of bosons, 
in agreement with the exact results. 

\section{Summary and conclusions}

In this contribution, we have studied the origin of the regular features 
that were obtained before in numerical studies of the IBM with random 
interactions. In particular, we addressed the dominance of $L=0$ ground 
states in the context of the vibron model and the IBM. In a mean-field 
analysis it was found that different regions of the parameter space can 
be associated with definite geometric shapes: a spherical shape, a deformed 
shape and a condensate of dipole (quadrupole) bosons for the vibron model 
(IBM). For both models, we obtained a good description of the probability 
distribution of ground state angular momenta. 

The studies with random interactions indicate 
that there is a significantly larger class of Hamiltonians that leads to 
regular, ordered behavior at the low excitation energies than was commonly 
assumed. The fact that these properties are shared by different models, 
seems to exclude an explanation solely in terms of the angular momentum 
algebra, the connectivity of the model space, or the many-body dynamics of 
the model, as has been suggested before. The present analysis points, at 
least for systems of interacting bosons, to a more general phenomenon that 
does not depend so much on the details of the angular momentum coupling, 
but rather on the occurrence of definite, robust geometric phases such as 
spherical and deformed shapes. For the nuclear shell model the situation 
is less clear. Although a large number of investigations to explain and 
further explore the properties of random nuclei have shed light on various 
aspects of the original problem, i.e. the dominance of $L=0$ ground states, 
in our opinion, no definite answer is yet available, and the full 
implications for nuclear structure physics are still to be clarified.  

In conclusion, the results presented in this article for the vibron model 
and the IBM may help to understand the origin of ordered spectra arising 
from random interactions, as has been observed in numerical studies of 
nuclear structure models.  

\section*{Acknowledgements}

This work was supported in part by CONACyT under projects 
32416-E and 32397-E, and by DPAGA under project IN106400.


\begin{thebibliography}{aa}

\bibitem{JBD}
C.W. Johnson, G.F. Bertsch and D.J. Dean,
Phys. Rev. Lett. {\bf 80}, 2749 (1998).

\bibitem{BF1}
R. Bijker and  A. Frank,
Phys. Rev. Lett. {\bf 84}, (2000), 420.

\bibitem{BF2} 
R. Bijker and A. Frank, 
Phys. Rev. C {\bf 62}, 014303 (2000).
 
\bibitem{BFP1}
R. Bijker, A. Frank and S. Pittel,
Phys. Rev. C {\bf 60}, 021302 (1999).

\bibitem{JBDT}
C.W. Johnson, G.F. Bertsch, D.J. Dean and I. Talmi,
Phys. Rev. C {\bf 61}, 014311 (2000).

\bibitem{BFP2}
R. Bijker, A. Frank and S. Pittel, 
Rev. Mex. F{\'{\i}}s. {\bf 46 S1}, 47 (2000). 

\bibitem{MVZ} 
D. Mulhall, A. Volya and V. Zelevinsky, 
Phys. Rev. Lett. {\bf 85}, 4016 (2000). 

\bibitem{DD}
D. Dean, 
Nucl. Phys. A {\bf 682}, 194c (2001). 

\bibitem{DW}
S. Dro\v{z}d\v{z} and M. W\'ojcik, 
preprint nucl-th/0007045.

\bibitem{ZA}
Y.M. Zhao and A. Arima, 
Phys. Rev. C {\bf 64}, 041301 (2001). 

\bibitem{Zuker}
V. Vel\'azquez and A.P. Zuker, 
preprint nucl-th/0106020. 

\bibitem{KZC}
D. Kusnezov, N.V. Zamfir and R.F. Casten, 
Phys. Rev. Lett. {\bf 85}, 1396 (2000). 

\bibitem{DK}
D. Kusnezov, 
Phys. Rev. Lett. {\bf 85}, 3773 (2000); \\
R. Bijker and A. Frank, 
Phys. Rev. Lett {\bf 87}, 029201 (2001); \\
D. Kusnezov, 
Phys. Rev. Lett {\bf 87}, 029202 (2001). 

\bibitem{BF3} 
R. Bijker and A. Frank, 
Phys. Rev. C {\bf 64}, 061303 (2001). 

\bibitem{vibron}
F. Iachello and R.D. Levine, 
{\it Algebraic theory of molecules} 
(Oxford University Press, Oxford, 1995). 

\bibitem{onno}
O.S. van Roosmalen and A.E.L. Dieperink, 
Ann. Phys. (N.Y.) {\bf 139}, 198 (1982); \\
O.S. van Roosmalen, Ph.D. thesis, Univ. of Groningen (1982). 

\bibitem{duke}
J. Dukelsky, G.G. Dussel, R.P.J. Perazzo, S.L. Reich and H.M. Sofia, 
Nucl. Phys. A {\bf 425}, 93 (1984). 

\bibitem{IBM}
F. Iachello and A. Arima,
{\it The interacting boson model}
(Cambridge University Press, Cambridge, 1987).

\bibitem{cs}
J.N. Ginocchio and M. Kirson, 
Phys. Rev. Lett. {\bf 44}, 1744 (1980); 
A.E.L. Dieperink, O. Scholten and F. Iachello, 
Phys. Rev. Lett. {\bf 44}, 1747 (1980). 

\end{thebibliography}
\end{document}